\documentclass[a4paper]{PoS}
\usepackage{lineno}
\title{On the TeV spectral energy distribution of HAWC's sources}

\ShortTitle{TeV SED of HAWC's sources}

\author{\speaker{Filiberto Hueyotl-Zahuantitla}\thanks{Commissioned to the FCFM-UNACH} $^{a}$ , C\'esar \'Alvarez$^b$, Roberto Arceo$^b$ and Karen Salom\'e Caballero$^b$\\
        \llap{$^a$}C\'atedra CONACYT, CONACYT-M\'exico\\ 
        \llap{$^b$}Universidad Aut\'onoma de Chiapas (FCFM-UNACH), Chiapas, M\'exico\\
        E-mail: \email{fhueyotl@conacyt.mx}, \email{crabpulsar@hotmail.com}, \email{roberto.arceo@unach.mx}, \email{karen.mora@unach.mx}}
\author{on behalf of the HAWC Collaboration}

\abstract{ We present a systematic study of the TeV gamma-ray spectrum
for the sources in the 2HWC HAWC catalog. Three spectrum models are considered in the analysis: simple power law, power law with exponential cut-off and a log-parabola. A comparison of the test statistic of nested models was performed in order to look for the best description. We used gamma-ray sky maps of 17 and 25 months of HAWC data.  From the analysis, only three sources of the catalog deviate from simple power law spectrum with high significance. The corresponding parameters are given in each case.}

\FullConference{35th International Cosmic Ray Conference --- ICRC2017\\
		10--20 July, 2017\\
		Bexco, Busan, Korea}

\begin{document}

\section{Introduction}
The spectral energy distribution (SED) provides information on the emission mechanisms powering astrophysical objects, traditionally classified as either leptonic or hadronic. 
In the TeV range the emission is often interpreted as due to Inverse Compton scattering , i.e. leptonic, leading to simple power law approximations to first order $\sim E^{-\alpha}$, where $E$ is the energy of the photon and $\alpha$ is the spectral index. However, there is strong evidence of deviations from simple power law and thus more complicated fits has been suggested: power law with exponential cut-off, log-parabola, broken power laws, and  others. Recent works suggest a log-parabola as the TeV spectrum model for the Crab Nebula, see for example \cite{1,2}. The TeV gamma-ray spectrum of the two blazars, Markarian 421 and Markarian 501 are modeled as power laws with exponential cut-off, see for instance \cite{3,4,8}. Here we present a comparative study of spectral fits of the sources given in the second HAWC catalog beyond the simple power law reported in \cite{5}.

\section{HAWC sample}
The High Altitude Water Cherenkov Observatory (HAWC) is located in central M\'exico at an elevation of 4100 meters a.s.l. HAWC consist of 300 water tanks with a footprint of 22,000 m$^2$. Each tank of 4.5 m high and 7.3 m diameter  is instrumented with three 8-in PMTs arrayed around  one central  10-in PMT at the bottom of the tank. The PMTs detect the Cherenkov light that relativistic particles emits as they cross the 200,000 L of  water contained in each tank. HAWC is the most sensitive wide field-of-view TeV telescope, with an instantaneous field of view of 2  steradians and > 95\% duty cycle, it continuously surveys and monitors the sky for gamma ray energies between 100 GeV to 100 TeV.
In this work we consider the point sources listed in Table 2 of the 2HWC catalog \cite{5}. From the list, two sources have been identified as extragalactic and are associated with blazars, seven may be pulsar wind nebulae (PWN), two supernova remnants(SNR), 14 have possible associations with PWN, SNR and molecular clouds. The remaining are unassociated.

\section{Method}
We used the so called Likelihood Fitting Framework (LiFF) \cite{6} to compute a test statistic (TS)
from HAWC maps. Two maps from HAWC data were used, for 17 and 25 months, both produced with a HEALPix grid of resolution of 1024. The TS is defined in term of the maximum likelihood ratio: 
\begin{equation}
TS=2 \ln \frac{\mathsf{L}^{\rm max}(\rm{source~model})}{\mathsf{L}(\rm{null~model})},
\end{equation}
where $\mathsf{L}^{\rm max}(\rm{source ~model})$ is the  likelihood that a source is present and $\mathsf{L}(\rm{null~model})$ is the null hypotheses that the observed events is due to background alone. The likelihood of a model $\mathsf{L}(\rm{model})$ is obtained by comparing the observed event counts with the expected counts, for all the pixels in the region of interest and for all energy bins. We used nine energy bins as in \cite{5} and \cite{8}. For the null model the expected counts are simply given by the background maps derived from data. For the source model the expected counts corresponds to the same background plus a signal contribution from the source derived from simulation. The physical model assume a point-like source and three energy spectrum laws for each source: simple power law, power law with exponential decay and a log-parabola. These models are contained in the following formula:
\begin{equation}
dN/dE = f_{0}(E/E_{0})^{-\alpha+\beta{\log{(E/E_{0})}}}\cdot \exp{(-E/E_{c})},
\end{equation}
where $dN$ is the flux of gamma photons in the energy interval $dE$, $f_{0}$ is the normalization flux at the reference energy $E_{0}=7$ TeV, the hardness of the
spectrum is given by the spectral index $\alpha$, $\beta$ is a second index known as the curvature parameter and $E_{c}$ is the cut-off energy. For $\beta =0$ and $E_{c}\rightarrow \infty$ we recover the simple power law, when $\beta=0$ we recover the cut-off power law, and when $E_{c}\rightarrow \infty$ we recover the log-parabola. The simple power law and the power law with exponential decay are nested, also the simple power law and the log-parabola. However, the power law with exponential cut-off and the log-parabola models are not nested. Therefore the last two models can not be compared each other by means of statistical test.  

With LiFF the TS is maximized with respect to the parameters of the spectrum model. We leave the respective parameters free for each case. We use a threshold $\Delta$TS > 25 between the nested models in order to see whether the observed spectrum deviates from the simple power law.

\begin{table}[!htbp]
\small{
\centering 
\begin{tabular}{llrlccc} % Es posible alinear toda la tabla adicionando opciones  
\hline
\hline            
Name          &Model&  TS      &$\Delta$TS&$f_{0}$              &$\alpha$   & $E_{c}$, $\beta$\\
              &     &          &          &$[\rm{TeV^{-1}~cm^{-2}s^{-1}}]$&           &                 \\
\hline\\

\multicolumn{7}{c}{17 Months of HAWC data}\\
2HWC J0534+220&SP &	11120.6& &1.84E-13$ \pm$	2.37E-15& 2.59 $\pm$	0.01&               \\
	      &CP &11249.3&128.7 &2.98E-13$ \pm$	1.77E-14& 2.25 $\pm$	0.04&	32.3 $\pm$ 4.9  \\
	      &LP &11262.0&141.4 &2.40E-13$ \pm$	6.19E-15& 2.64 $\pm$	0.02&	0.15 $\pm$ 0.016\\		
2HWC J1104+381&SP &1315.0&	 &7.09E-14$ \pm$	2.86E-15& 3.04 $\pm$	0.03&	                \\
	      &CP &1385.8&70.8 &3.52E-13$ \pm$	1.56E-13& 2.27 $\pm$	0.22&	5.7 $\pm$ 1.9  \\
	      &LP &1389.7&74.7 &8.82E-14$ \pm$	5.50E-15& 3.61 $\pm$	0.14&	0.35 $\pm$ 0.071\\	
2HWC J1653+397&SP &572.7&	 &5.66E-14$ \pm$	2.70E-15& 2.86 $\pm$	0.04&		        \\
	      &CP &623.6&50.9 &3.84E-13$ \pm$	2.62E-13& 1.61 $\pm$	0.48&	5.6 $\pm$ 2.5  \\
	      &LP &623.7&60.0 &1.01E-13$ \pm$	1.01E-14& 3.14 $\pm$	0.13&	0.51 $\pm$ 0.151\\
							
\hline\\
\multicolumn{7}{c}{25 Months of HAWC data}\\		
2HWC J0534+220&SP &17189.1& &1.88E-13$ \pm$	2.0E-15&  2.58 $\pm$	0.01&                   \\
	      &CP &17409.1&220.0 &3.17E-13$ \pm$	1.57E-14& 2.23 $\pm$	0.04&	29.4 $\pm$ 3.45 \\
              &LP &17423.5&234.4 &2.48E-13$ \pm$	5.23E-15& 2.64 $\pm$	0.01&	0.155 $\pm$ 0.014\\		
2HWC J1104+381&SP &1523.8&	 &6.27E-14$ \pm$	2.3E-15&  3.03 $\pm$	0.02&                    \\
              &CP &1625.3&101.6 &4.32E-13$ \pm$	1.75E-13& 2.10 $\pm$	0.20&	4.6 $\pm$ 1.19 \\
              &LP &1630.0&106.3 &7.95E-14$ \pm$	4.80E-15& 3.70 $\pm$	0.15&	0.412 $\pm$ 0.077\\		
2HWC J1653+397&SP &474.0&	 &4.12E-14$ \pm$	2.1E-15&  2.88 $\pm$	0.04&		         \\
              &CP &515.5&41.5 &2.37E-13$ \pm$	1.75E-13& 1.75 $\pm$	0.51&	6.3 $\pm$ 3.30 \\
              &LP &515.2&41.2 &7.18E-14$ \pm$	7.60E-15& 3.13 $\pm$	0.13&	0.472 $\pm$ 0.149\\

\hline
\end{tabular}
\caption{TeV gamma-ray spectrum models and the corresponding parameters. In the first column, from the top to the bottom, the sources correspond to: Crab Nebula, Markarian 421 and Markarian 501. The parameters are described in the text. The uncertainties in the parameters are statistical only.  Discussion on the systematic uncertainties are given in \cite{1}.} 
}
\label{table1}
\end{table}

%-------------------------------------------------------------------------------------
\begin{table}[!htbp]
\small{
\centering 
\begin{tabular}{lrccc} % Es posible alinear toda la tabla adicionando opciones  
\hline
\hline            
Name          &TS   &  $\alpha$&$f_{0}$                        &  TeVCat\\
              &     &          &$[\rm{TeV^{-1}~cm^{-2}s^{-1}}]$&           \\
\hline\\
					
2HWC J0631+169&	31.7&	2.73$\pm$	0.14&	6.13E-15$\pm$	1.15E-15&Geminga \\
2HWC J0635+180&	31.9&	2.57$\pm$	0.15&	5.77E-15$\pm$	1.23E-15&Geminga \\
2HWC J1809-190&	126.3&	2.64$\pm$	0.09&	8.34E-14$\pm$	1.22E-14&HESS J1809-193 \\
2HWC J1814-173&	230.9&	2.59$\pm$	0.07&	8.99E-14$\pm$	1.07E-14&HESS J1813-178 \\
2HWC J1819-150&	80.4&	2.83$\pm$	0.10&	5.14E-14$\pm$	6.63E-15&SNR G015.4+00.1 \\
2HWC J1825-134&	1227.2&	2.57$\pm$	0.03&	1.42E-13$\pm$	6.63E-15&HESS J1826-130 \\
2HWC J1831-098&	143.8&	2.79$\pm$	0.08&	4.09E-14$\pm$	3.87E-15&HESS J1831-098 \\
2HWC J1837-065&	918.5&	2.87$\pm$	0.03&	8.85E-14$\pm$	3.29E-15&HESS J1837-069 \\
2HWC J1844-032&	459.8&	2.65$\pm$	0.05&	4.64E-14$\pm$	2.59E-15&HESS J1844-030 \\
2HWC J1847-018&	222.4&	2.85$\pm$	0.06&	3.02E-14$\pm$	2.19E-15&HESS J1848-018 \\
2HWC J1849+001&	183.0&	2.50$\pm$	0.08&	2.06E-14$\pm$	2.33E-15&IGR J18490-0000 \\
2HWC J1852+013&	83.8&	2.89$\pm$	0.10&	1.57E-14$\pm$	1.86E-15&- \\
2HWC J1857+027&	452.7&	2.94$\pm$	0.04&	3.44E-14$\pm$	2.01E-15&HESS J1857+026 \\
2HWC J1902+048&	48.8&	3.16$\pm$	0.16&	7.70E-15$\pm$	1.95E-15&- \\
2HWC J1907+084&	69.0&	3.24$\pm$	0.13&	7.38E-15$\pm$	1.85E-15&- \\
2HWC J1908+063&	646.6&	2.58$\pm$	0.04&	3.79E-14$\pm$	1.79E-15&MGRO J1908+06 \\
2HWC J1912+099&	120.6&	3.01$\pm$	0.08&	1.30E-14$\pm$	1.57E-15&HESS J1912+101 \\
2HWC J1914+117&	55.0&	2.97$\pm$	0.12&	8.40E-15$\pm$	1.42E-15&- \\
2HWC J1921+131&	30.0&	2.72$\pm$	0.16&	6.29E-15$\pm$	1.22E-15&- \\
2HWC J1922+140&	87.4&	2.65$\pm$	0.10&	1.04E-14$\pm$	1.27E-15&W 51 \\
2HWC J1928+177&	99.0&	2.57$\pm$	0.10&	1.02E-14$\pm$	1.32E-15&- \\
2HWC J1930+188&	72.8&	2.78$\pm$	0.10&	9.60E-15$\pm$	1.20E-15&SNR G054.1+00.3 \\
2HWC J1938+238&	35.8&	3.12$\pm$	0.15&	5.40E-15$\pm$	1.41E-15&- \\
2HWC J1953+294&	37.6&	2.77$\pm$	0.13&	7.57E-15$\pm$	1.29E-15&- \\
2HWC J1955+285&	26.7&	2.67$\pm$	0.19&	6.06E-15$\pm$	1.32E-15&- \\
2HWC J2006+341&	49.3&	2.80$\pm$	0.12&	9.74E-15$\pm$	1.46E-15&- \\
2HWC J2019+367&	646.5&	2.36$\pm$	0.05&	3.27E-14$\pm$	2.46E-15&VER J2019+368 \\
2HWC J2020+403&	90.6&	3.07$\pm$	0.09&	1.65E-14$\pm$	2.26E-15&VER J2019+407 \\
2HWC J2024+417&	29.7&	2.69$\pm$	0.18&	1.01E-14$\pm$	2.17E-15&MGRO J2031+41 \\
2HWC J2031+415&	311.5&	2.52$\pm$	0.06&	3.03E-14$\pm$	2.67E-15&TeV J2032+4130 \\

\hline
\end{tabular}
\caption{2HWC sources with $\Delta$TS<25. The columns correspond to: HAWC source name, test statistic, spectral index, normalization flux at 7 TeV and
the nearest TeVCat source name. The uncertainties are statistical only. The results were obtained with the map of 25 months.} 
}
\label{table1}
\end{table}

\section{Results}

 We compared the TS between nested models, namely, simple power law (SP) versus power law with exponential cut-off (CP) and simple power law versus log-parabola (LP). Table 1 presents the sources of the catalog that fulfill the condition $\Delta$TS>25, these sources are 2HWC J0534+220 (Crab Nebula), 2HWC J1104+381 (Markarian 421) and 2HWC J1653+397 (Markarian 501). The upper and the lower halves of the table are based on the 17 months and 25 months of data, respectively.  
The HAWC name of each source is given in column 1, column 2 gives the assumed spectrum model, TS is in column 3, $\Delta$TS is given in column 4, column 5 shows the normalization flux $f_{0}$  at the reference energy $E_{0}=$ 7 TeV, the spectral index $\alpha$ appears in column 6, column 7 presents the cut-off energy $E_{c}$ or the curvature parameter $\beta$ depending on the spectrum model given in column 2. Note that $\Delta$TS is slightly higher for SP versus LP than SP versus CP for the three sources in the case of 17 months of data. In the case of 25 months of HAWC data, $\Delta$TS is slightly lower for SP versus LP than SP versus CP for Markarian 501. However we can not conclude which one of LP or CP is the most appropriate model for the spectrum because these are not nested models and therefore can not be compared directly via TS. 
The results presented here are consistent with those presented by \cite{8} in the case of Markarians and by \cite{1} for the Crab Nebula. 

In Table 2 we present the sources with $\Delta$TS<25, according to this condition the spectrum of these sources are better described by a simple power law fit, in line with \cite{5}. The parameters of the spectrum are given for each source. The map of 25 months was used in this case.

\section{Conclusions}
We studied the TeV gamma-ray spectra of the sources in the 2HWC catalog by applying different models to the spectra. 
We conclude that the TeV spectra for the Crab Nebula, Markarian 421 and Markarian 501 deviate from a simple power law.  However we can not disentangle whether a power law with exponential cut-off or log-parabola is the most appropriate description for the TeV spectrum since they are not nested models and thus can not be compared via the TS. A detailed comparison between power law with exponential cut-off and log-parabola will be addressed in a forthcoming communication. The simple power law is a better
description to most spectra of the HAWC catalog sources, this results is in line with the
2HWC catalog publication.\\ 

We acknowledge the support from: the US National Science Foundation (NSF); the 
US Department of Energy Office of High
-Energy Physics; the Laboratory Directed Research and Development (LDRD) program of Los Alamos National Laboratory; Consejo Nacional de Ciencia y Tecnolog\'{\i}a (CONACyT), M{\'e}xico (grants 271051, 232656, 260378, 179588, 239762, 254964, 271737, 258865, 243290, 132197), Laboratorio Nacional HAWC de rayos gamma; Red HAWC, M{\'e}xico; DGAPA-UNAM (grants RG100414, IN111315, IN111716-3, IA102715, 109916, IA102917); VIEP-BUAP; the University of Wisconsin Alumni Research Foundation; the Institute of Geophysics, Planetary Physics, and Signatures at Los Alamos National Laboratory; Polish Science Cen
tre grant DEC-2014/13/B/ST9/945; Coordinaci{\'o}n de la Investigaci{\'o}n Cient\'{\i}fica de la Universidad Michoacana; FCFM-UNACH PFCE 2016.  Thanks to Luciano D\'{\i}az and Eduardo Murrieta for technical support. FHZ wishes to thank to CONACYT, C\'atedra CONACYT 1563.


\begin{thebibliography}{99}
\bibitem{1}
Abeysekara et al., {\it Observations of the Crab Nebula with the HAWC Gamma-ray Observatory,}ApJ 843 (June 2017) 39, ArXiv e-prints (Jan., 2017) [{\ttfamily arXiv:1701.01778}]
\bibitem{2}
F. Fraschetti and M. Pohl, {\it Particle acceleration model for the broadband baseline spectrum of the Crab nebula,} (Feb., 2017) [{\ttfamily arXiv:1702.00816}]
\bibitem{3}
S. Sahu, L. S. Miranda and S. Rajpoot, Mrk421: {\it The Multi-TeV emission and its astrophysical origin,} J. Phys: Conf Ser. 761 (2016) 012042
\bibitem{4}
D. Dorner et al., {\it First Study of combined Blazar light curves with FACT and HAWC,} ArXiv e-prints (Oct., 2016) [{\ttfamily arXiv:1610.06627}]
\bibitem{5}
Abeysekara et al., {\it The 2HWC HAWC Observatory Gamma Ray Catalog,} ArXiv e-prints (Feb., 2017) [{\ttfamily arXiv:1702.02992}]
\bibitem{6}
P. W. Younk, R. J. Lauer, G. Vianello, et al., \emph{A high-level analysis framework for HAWC}, in proceedings of the \emph{34th International Cosmic Ray Conference}, \pos{PoS(ICRC2015)948} (2015).
\bibitem{7}
Q. Zhu, D. Yan, P. Zhang et al., {\it Testing one-zone synchrotron-self-Compton models with spectral energy distributions of Mrk 421}, MNRAS 463 (Dec., 2016) 4481 [{\ttfamily arXiv:1609.04170}]
\bibitem{8}
A. U. Abeysekara et al.,  {\it Dayly Monitoring of TeV Gamma-Ray Emission from Mrk 421, Mkr 501, and the Crab Nebula with HAWC}, ApJ, 841 (May 2017) 100 [{\ttfamily arXiv:1703.06968}].

\end{thebibliography}
\end{document}